\documentclass[%
 reprint,
 amsmath,amssymb,
 aps, doi= false,
]{revtex4-2}
\usepackage{graphicx}
\usepackage{dcolumn}
\usepackage{bm}
\usepackage{natbib}

\usepackage{enumerate}
\usepackage{enumitem}
\usepackage{hyperref}
\usepackage{cleveref}
\usepackage{graphicx}
\newcommand{\bs}[1]{\boldsymbol{#1}}

\newcommand{\rint}[1]{\int \frac{\mathrm{d}^4{#1}}{(2\pi)^4}}
\begin{document}

\preprint{APS/123-QED}

\title{Quantum corrections to tunnelling amplitudes of neutral scalar fields}

\author{Rosemary Zielinski$^{1}$}
\email{rosemary.zielinski@anu.edu.au}
\author{C\'edric Simenel$^{1,2}$}%
\email{cedric.simenel@anu.edu.au}
\author{Patrick McGlynn$^{1,2}$}
\email{Current address: Facility for Rare Isotope Beams, Michigan State University, East Lansing, Michigan 48824, USA}

\affiliation{%
 $^{1}$Department of Fundamental and Theoretical Physics, The Australian National University
}%

\affiliation{%
 $^{2}$Department of Nuclear Physics and Accelerator Applications, The Australian National University
}%

\date{\today}

\begin{abstract}
Though theoretical treatments of quantum tunnelling within single-particle quantum mechanics are well-established, at present, there is no quantum field-theoretic description (QFT) of tunnelling. Due to the single-particle nature of quantum mechanics, many-particle effects arising from quantum field theory are not accounted for. Such many-particle effects, including pair-production, have proved to be essential in resolving the Klein-paradox.
This work seeks to determine how quantum corrections affect the tunnelling probability through an external field. We investigate a massive neutral scalar field, which interacts with an external field in accordance with relativistic quantum mechanics.  To consider QFT corrections, we include another massive quantised neutral scalar field coupling to the original via a cubic interaction. This study formulates an all-order recursive expression for the loop-corrected scalar propagator, which contains only the class of vertex-corrected Feynman diagrams. This equation applies for general external potentials. Though there is no closed-form analytic solution, we also demonstrate how to approximate the QFT corrections if a perturbative coupling to the quantised field is assumed. 
\end{abstract}

\maketitle
Notwithstanding the extensive application of quantum tunnelling, ranging from tunnel diodes \cite{Esaki_1958}, 
scanning tunnelling microscopy \cite{binnig_scanning_1987, binnig_scanning_1983}, 
and its importance to many biological \cite{Devault_1980,trixler_quantum_2013} 
and chemical systems, fundamental theoretical questions about tunnelling remain unanswered. Even simple questions, such as the time a particle takes to tunnel, are still disputed \cite{ramos_measurement_2020,Hartman_1962,gavassino_subluminality_2023,dumont_relativistic_2020}. 
Quantum tunnelling through external potentials is typically understood in the framework of single-particle quantum mechanics using the non-relativistic Schr\"odinger equation or the relativistic Klein-Gordon and Dirac equations. However, these descriptions fail to account for particle number non-conserving processes. Yet, there is no comprehensive description of quantum tunnelling also compatible with quantum field theory (QFT). Because interactions in a QFT framework are described by local couplings to mediator fields, all interactions involve the destruction and creation of virtual particles naturally resulting in a many-particle theory. Many-particle effects including virtual particle mediators and pair production have been shown to have important measurable consequences even in low-energy systems. Notably, QFT predicts the anomalous electron magnetic moment \cite{schwinger_quantum-electrodynamics_1948,aoyama_revised_2018} through a quantised photon field. Hyperfine electronic structure, such as the Lamb shift \cite{lamb_fine_1947,bethe_electromagnetic_1947}, is the result of electron self-energy corrections, vertex corrections, and vacuum polarization contributions (i.e. the Uehling potential \cite{uehling_1935}), all of which arise from the quantization of electromagnetic and fermionic fields. 

In the context of quantum tunnelling, these many-particle effects manifest in the Klein Paradox \cite{Klein_1929}, where fermions incident on a step potential above the Schwinger limit ($eV>2m$) display violations of unitarity. The paradox is resolved by including pair-production at the barrier  \cite{hansen_kleins_1981}, an intrinsically QFT effect going beyond the physics of relativistic quantum mechanics. Other work has considered the possibility of `tunnelling of the third kind' \cite{gies_2009,gardiner_2013} whereby a particle interacting with a barrier may split into a pair of virtual particles which interact only weakly with the barrier, and recombine after a finite distance. Additionally, while quantum tunnelling between field configurations has been well-studied using instanton methods \cite{coleman_uses_1979} and applied to false vacuum decay hypotheses \cite{coleman_fate_1977,devoto_false_2022,devoto_false_2022}
this is conceptually distinct from our work which considers quantum tunnelling of a \textit{particle} through external localised potentials. 

Merging QFT with a theory of quantum tunnelling is difficult due to the perturbative formalisms which dominate QFT calculations. Many scattering calculations employ the scattering matrix ($S$-matrix), a unitary time evolution operator with the expression 
\begin{align}
S &= \mathcal{T}\left [\mathrm{e}^{-i\int \mathrm{d}^4x H_{int}(x)}\right ]. 
\end{align}
Here $\mathcal{T}$ denotes the time-ordering operator, and $H_{int}(x)$ the interaction Hamiltonian. There are no analytic solutions for the complete interacting $S$-matrix in four-dimensions for any non-trivial QFT. In practice, the $S$-matrix is computed perturbatively via a truncation of the following series 
\begin{widetext}
\small
\begin{align}
S = \sum _{n=0}^{\infty} \frac{(-i)^n}{n!}\int \mathrm{d}^4x_1\int \mathrm{d}^4x_2 \dots \int \mathrm{d}^4x_n \mathcal{T}\left [H_{int}(x_1)H_{int}(x_2)\dots H_{int}(x_n)\right ],
\end{align}
\end{widetext}
where each order in $H_{int}(x)$ can be represented by a set of Feynman diagrams. Such an approach is fundamentally incompatible with quantum tunnelling, which is a non-perturbative phenomenon (in the interaction Hamiltonian). 
This warrants an alternative approach to integrating QFT with quantum tunnelling. 

Previous work has described electron scattering from an external potential, using both the canonical quantisation \cite{de_leo_2009} and path integral formalism
\cite{Xu_2016} of QFT. However, both studies employed the single-particle relativistic Dirac equation and neglected a quantised photon field, treating the external field classically. While QFT methods were used, the underlying physics is restricted to relativistic quantum mechanics. Additionally, the single-particle scattering calculations presented in \cite{de_leo_2009,Xu_2016} were exclusively above-barrier, and therefore could not describe tunnelling. Our recent work \cite{zielinski_2024_eprint} built upon their formalism using a simpler model of a neutral scalar field interacting with an external mass perturbation-like potential. We demonstrated that for simple delta function potentials, tunnelling amplitudes could be obtained via analytic continuation of the $S$-matrix and an infinite sum of Feynman diagrams. Though this was successful in recovering tunnelling amplitudes consistent with RQM, it too was limited by the single-particle nature of the Klein-Gordon equation. The present work seeks to extend both our previous study and \cite{de_leo_2009, Xu_2016} to include many-particle effects from an additional quantised field. Our model describes the dynamics of a neutral scalar field interacting both with an external scalar field (acting as an external potential) and a quantised scalar field. A key result is the derivation of a scalar propagator which accounts for all-order interactions with the scalar field and the external field, via a summation of a restricted class of Feynman diagrams. This process is conceptually similar to the self-consistent Hartree-Fock diagrammatic expansion \cite{fetter_2003}. We also present a perturbative coupling approximation to the dressed propagator, which still treats the external field to all orders.  Though we derive a dressed propagator, the extraction of tunnelling amplitudes remains challenging due to the complexity of the integral equation, and is left for future work. However, this work provides the necessary formalism to integrate QFT corrections to tunnelling, and therefore to consider more physically interesting Lagrangians. 

\subsection*{Formalism}
The primary goal of this work is to develop a formalism that accounts for quantum corrections which affect tunnelling through external localised barriers. In order to do this, we consider the simplest system for which these effects may be probed. To this end, this work considers a massive neutral scalar field $\phi$, interacting with an external field $u(x)$, and another massive neutral scalar field $\Phi$. The associated Lagrangian is:
\small
\begin{align}
\mathcal{L} &= \frac{1}{2}(\partial_{\nu}\Phi)^2-\frac{1}{2}m'^2\Phi^2+\frac{1}{2}(\partial_{\nu}\phi)^2\nonumber\\
&-\frac{1}{2}m^2\phi^2-\frac{1}{2}eu(x)\phi^2-\frac{\mu}{2}\Phi\phi^2, \label{eqn:full_lagrangian}
\end{align} 
\normalsize
where the neutral scalar fields have a cubic interaction via the last term. Note that the scalar field $\Phi$, does not interact with the external field $u(x)$, and that the external field, by definition, has no dynamical term in the Lagrangian. Without the addition of the dynamical field $\Phi$, the Lagrangian would be equivalent to a single-particle theory described by the Klein-Gordon Lagrangian with a mass perturbation.

 The choice of cubic interaction is somewhat arbitrary --- in principle, any other renormalizable interaction would allow an investigation into many-particle effects. However, the cubic term is one of the simpler choices: it produces a super-renormalizable theory, and also includes vertex-corrections to the theory at order $\mu^2$. The effect of the field $\Phi$ is to add an interaction vertex,
\begin{align}
-i\mu = \vcenter{\hbox{\includegraphics[scale=0.7]{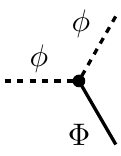}}},
\end{align}
in addition to the interaction with the external field, 
\begin{align}
-ie\tilde{u}(p-k) &= \vcenter{\hbox{\includegraphics[scale=0.7]{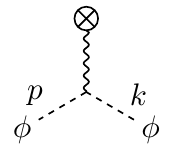}}}
\end{align}
where $\otimes$ denotes the external field, and $\tilde{u}(p-k)$ the Fourier transform of the external potential. 

In the context of tunnelling, we are concerned with single-particle incoming and outgoing states of the field $\phi$, characterised with initial momentum $p = (p_0, 0 ,0, p_3)$ and final $k = (k_0, k_1, k_2, k_3)$ respectively. To obtain tunnelling amplitudes, the complete interacting $S$-matrix element $\langle k | S | p\rangle$ must be determined. For a time-independent potential $u(x)$, which is only a function of $x_3$, the transmission and reflection amplitudes are related to the $S$-matrix via \cite{Xu_2016,de_leo_2009}
\begin{align}
 T =& \int_{-\infty}^{\infty}\mathrm{d}k_1\int_{-\infty}^{\infty}\mathrm{d}k_2\int_0^{\infty}\frac{\mathrm{d}k_3}{(2\pi)^32E(\bs{k})}\langle k | S | p\rangle \\
 R =& \int_{-\infty}^{\infty}\mathrm{d}k_1\int_{-\infty}^{\infty}\mathrm{d}k_2\int_{-\infty}^{0}\frac{\mathrm{d}k_3}{(2\pi)^32E(\bs{k})}\langle k | S | p\rangle.
\end{align}

We have the diagrammatic representation of the $S$-matrix element, 
\begin{equation}
\langle k |S |p \rangle = \vcenter{\hbox{\includegraphics[scale=1.0]{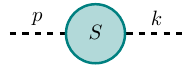}}},
\end{equation}
where the circle represents the exact, all-order interacting $S$-matrix. For calculational purposes, we consider instead the two-point correlation function $G(q,q')$, diagrammatically related to the $S$-matrix via
\begin{equation}
G(q,q') = \vcenter{\hbox{\includegraphics[scale=1.0]{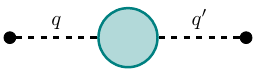}}} =\vcenter{\hbox{ \includegraphics[scale=1.0]{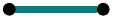}}}. \label{eqn:full_interacting_diagram}
\end{equation}
This is simply a dressed propagator, where the legs of the diagram are no longer required to be on mass-shell. More formally, $S$-matrix elements can be directly extracted from two-point correlation functions via the LSZ reduction formula \cite{lehmann_zur_1955}. Note also that momentum is not necessarily conserved --- by definition, the presence of an external field requires this. Within this propagator formalism, the Feynman rules are as follows:
\begin{enumerate}
\item Each $\phi^2u(x)$ vertex, with incoming scalar momentum $p$ and outgoing scalar momentum $k$ has a vertex factor of $-ie\tilde{u}(p-k)$ (links a wavy line to a dashed line). 
\item Each $\phi^2\Phi$ vertex has a factor of $-i\mu$.
\item Internal scalar ($\phi$) lines have a scalar propagator $D(q,m) = \frac{i}{q^2-m^2+i\varepsilon}$ (dashed lines), while internal scalar ($\Phi$) lines have an associated propagator $D(q,m') = \frac{i}{q^2-m'^2+i\varepsilon}$ (solid lines). 
\item A dressed line (thick solid line), with incoming momenta $p$ and outgoing momenta $k$, contributes a factor of $G(p,k)$.
\item All unspecified momenta of internal lines are individually integrated over with measure $\int\frac{\mathrm{d}^4q}{(2\pi)^4}$.
\end{enumerate} 
It is well-known that there are no exact solutions of interacting theories in four dimensions within QFT, such that a closed form expression for this particular dressed propagator is not known. It therefore remains to find a suitable approximation which can generate meaningful quantum corrections to quantum tunnelling, without requiring a solution to the fully-interacting theory. To do this, we take inspiration from the diagrammatic derivation of the self-consistent Hartree-Fock propagator \cite{fetter_2003}. 
\subsection*{Approximated dressed propagator \label{sec:dressed_propagator}}
\Cref{eqn:full_interacting_diagram} encodes an infinite number of Feynman diagrams, to all orders in $\mu$ and $u(x)$, with all topologies. One approximation to this dressed propagator restricts the class of Feynman diagrams, such that:
\begin{widetext}
\small
\begin{align}
\vcenter{\hbox{ \includegraphics[scale=0.6]{Figures/extra_feynman_diagrams_paper_2/solid_dressed_line.pdf}}} \approx & \vcenter{\hbox{\includegraphics[scale=0.6]{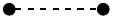}}} +\vcenter{\hbox{ \includegraphics[scale=0.6]{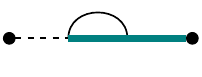}}}+\vcenter{\hbox{\includegraphics[scale=0.6]{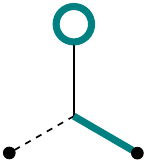}}} +\vcenter{\hbox{\includegraphics[scale=0.6]{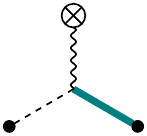}}} \label{eqn:dyson_self_consistent}\\
=& \underbrace{\vcenter{\hbox{\includegraphics[scale=0.6]{Figures/chapter_3/feynman_diagrams/free_scalar.pdf}}}+ \vcenter{\hbox{\includegraphics[scale=0.6]{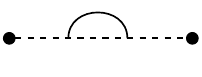}}}+\vcenter{\hbox{\includegraphics[scale=0.6]{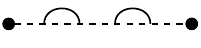}}}+\vcenter{\hbox{ \includegraphics[scale=0.6]{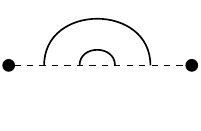}}}+ \vcenter{\hbox{ \includegraphics[scale=0.5]{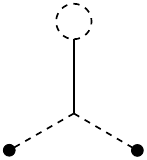}}}+ \vcenter{\hbox{ \includegraphics[scale=0.5]{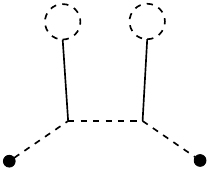}}}+ \vcenter{\hbox{ \includegraphics[scale=0.5]{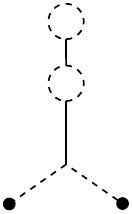}}} +  \vcenter{\hbox{ \includegraphics[scale=0.5]{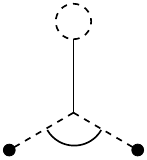}}}+\cdots}_{\text{self-energy contributions}}\nonumber 
\\
+& \underbrace{\vcenter{\hbox{\includegraphics[scale=0.6]{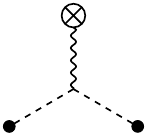}}} + \vcenter{\hbox{\includegraphics[scale=0.6]{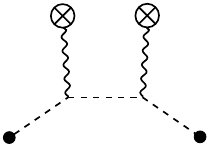}}}+\vcenter{\hbox{\includegraphics[scale=0.6]{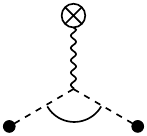}}}+\vcenter{\hbox{\includegraphics[scale=0.6]{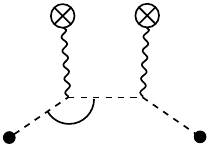}}}+\cdots}_{\text{vertex-corrected diagrams}}
\end{align}
\end{widetext}
\normalsize
This is analogous to the self-consistent Hartree-Fock approximation, where the second term in \cref{eqn:dyson_self_consistent} corresponds to the exchange term, while the third diagram to the direct term. We emphasise that the above equation is self-consistent because the dressed propagator is also used for any internal lines on the r.h.s. Additionally, \cref{eqn:dyson_self_consistent} is still non-perturbative in the couplings $\mu$ and $eu(x)$, and still describes an infinite number of Feynman diagrams. However, some diagrams present in \cref{eqn:full_interacting_diagram} are never generated in the self-consistent expansion. For example, the 1PI diagram
\begin{equation}
\vcenter{\hbox{\includegraphics[scale=0.6]{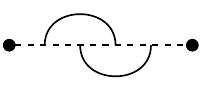}}}\label{eqn:absent_1PI}
\end{equation}
is included in \cref{eqn:full_interacting_diagram}, but is not generated by \cref{eqn:dyson_self_consistent}. Despite the approximations made, even this self-consistent dressed propagator is not tractable. We make a further simplification, and neglect the self-energy contributions to the propagator. Neglecting the self-energy terms does not account for the renormalised mass of the field $\phi$ (it should be noted the $\Phi$ self -energy terms are also absent). Such self-energy contributions will not affect the neutral scalar field interaction with the \textit{external} field, because self-energy type diagrams/subgraphs necessarily conserve momentum, given an $S$-matrix element between identical free single-particle states.  While transmitted particle have the same momentum as the incoming particles (for potentials which tend to zero for $x_3 \to \pm \infty$), the $S$-matrix elements also include the reflection contribution which does not conserve momentum asymptotically. Thus, any $S$-matrix element which perfectly conserves momentum cannot describe tunnelling, or a momentum-exchanging interaction with the field.  Crucially, because the self-energy contributions do not affect the interactions with the external field, they will also not be relevant for tunnelling. The self-consistent propagator can now be redefined, notated by a double-line:
\begin{widetext}
\begin{align}
\vcenter{\hbox{\includegraphics[scale=0.7]{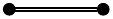}}} =\vcenter{\hbox{\includegraphics[scale=0.7]{Figures/chapter_3/feynman_diagrams/ext_interaction_1.pdf}}}+\vcenter{\hbox{\includegraphics[scale=0.7]{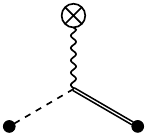}}}+\vcenter{\hbox{\includegraphics[scale=0.7]{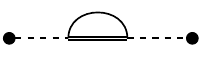}}}+\vcenter{\hbox{\includegraphics[scale=0.7]{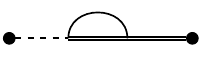}}}. \label{eqn:dyson_modified}
\end{align}
\end{widetext}
The implications of this approximation are that we only generate terms we call `vertex-corrected diagrams'. Additionally, this expression only includes the interacting terms in the propagator: there is no free-propagator generated in this expansion. Hence, subsequent tunnelling calculations must include the non-interacting contribution post-hoc. The omission of the free propagator in \cref{eqn:dyson_modified} is necessary to prevent the generation of self-energy terms, which we explicitly are removing. To consider the consequences in more detail, we define a new quantity, $G^{n}$, which is the $n$-th recursion of the vertex-corrected propagator into the equation above. Diagrammatically, 
\begin{widetext}
\begin{align}
\vcenter{\hbox{\includegraphics[scale=0.8]{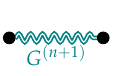}}} = \vcenter{\hbox{\includegraphics[scale=0.8]{Figures/chapter_3/feynman_diagrams/ext_interaction_1.pdf}}}+\underbrace{\vcenter{\hbox{\includegraphics[scale=0.8]{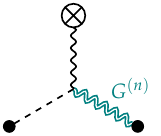}}}}_{a}+\underbrace{\vcenter{\hbox{\includegraphics[scale=0.8]{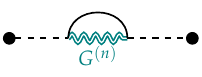}}}}_{b}+\underbrace{\vcenter{\hbox{\includegraphics[scale=0.8]{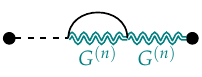}}}}_{c}, \label{eqn:nth_dyson}
\end{align}
\end{widetext}
Note that $G\neq \sum_n G^{(n)}$, rather, $G^{(\infty)} = G$. We define the zeroth `generation' to be: $
\vcenter{\hbox{\includegraphics[scale=0.75]{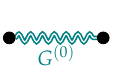}}} = \vcenter{\hbox{\includegraphics[scale=0.5]{Figures/chapter_3/feynman_diagrams/ext_interaction_1.pdf}}}$,
which is inserted into $G^{(1)}$ to yield:
\begin{widetext}
\begin{align}
\vcenter{\hbox{\includegraphics[scale=0.7]{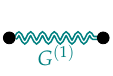}}} = \vcenter{\hbox{\includegraphics[scale=0.7]{Figures/chapter_3/feynman_diagrams/ext_interaction_1.pdf}}}+\vcenter{\hbox{\includegraphics[scale=0.7]{Figures/chapter_3/feynman_diagrams/double_external_vertex.pdf}}}+\vcenter{\hbox{\includegraphics[scale=0.7]{Figures/chapter_3/feynman_diagrams/vertex_correction.pdf}}}+\vcenter{\hbox{\includegraphics[scale=0.7]{Figures/chapter_3/feynman_diagrams/dressed_expansion/first/vertex_correction_leg.pdf}}},
\end{align}
\end{widetext}
where each additional term corresponds exactly to $G^0$ inserted into $a, b$ and $c$. Continuing to the second generation, we have: 
\begin{widetext}
\begin{align}
\vcenter{\hbox{\includegraphics[scale=0.8]{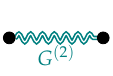}}} =&\vcenter{\hbox{\includegraphics[scale=0.4]{Figures/chapter_3/feynman_diagrams/ext_interaction_1.pdf}}}+\underbrace{ \vcenter{\hbox{\includegraphics[scale=0.4]{Figures/chapter_3/feynman_diagrams/double_external_vertex.pdf}}}+\vcenter{\hbox{\includegraphics[scale=0.4]{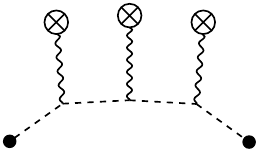}}}+\vcenter{\hbox{\includegraphics[scale=0.4]{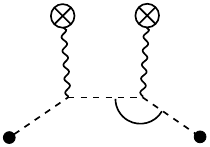}}}+\vcenter{\hbox{\includegraphics[scale=0.4]{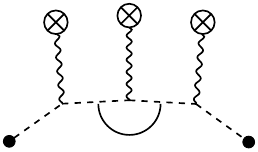}}}}_{\text{insertions into } a}\nonumber\\
&\quad\quad+ \underbrace{\vcenter{\hbox{\includegraphics[scale=0.4]{Figures/chapter_3/feynman_diagrams/vertex_correction.pdf}}}+\vcenter{\hbox{\includegraphics[scale=0.4]{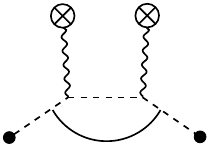}}}+\vcenter{\hbox{\includegraphics[scale=0.4]{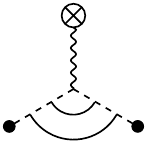}}}+\vcenter{\hbox{\includegraphics[scale=0.4]{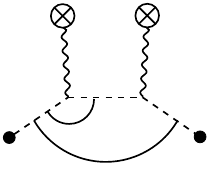}}}}_{\text{insertions into } b}\\
\substack{\text{insertions} \\ \text{into c}}&
\begin{cases}
&+ \vcenter{\hbox{\includegraphics[scale=0.4]{Figures/chapter_3/feynman_diagrams/dressed_expansion/first/vertex_correction_leg.pdf}}}+\vcenter{\hbox{\includegraphics[scale=0.4]{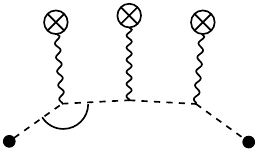}}}+\vcenter{\hbox{\includegraphics[scale=0.4]{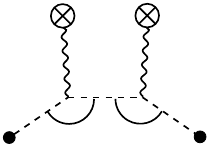}}}+\vcenter{\hbox{\includegraphics[scale=0.4]{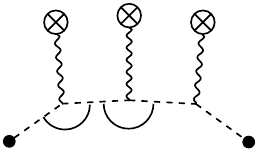}}}\nonumber\\
&+\vcenter{\hbox{\includegraphics[scale=0.4]{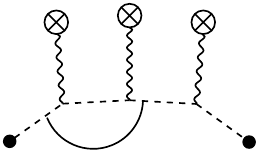}}}+\vcenter{\hbox{\includegraphics[scale=0.4]{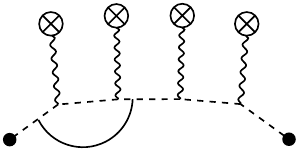}}}+\vcenter{\hbox{\includegraphics[scale=0.4]{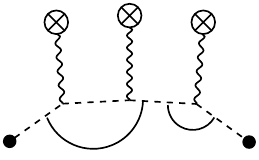}}}+\vcenter{\hbox{\includegraphics[scale=0.4]{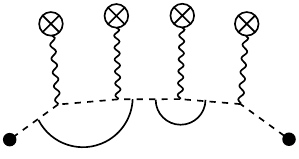}}}\nonumber\\
&+\vcenter{\hbox{\includegraphics[scale=0.4]{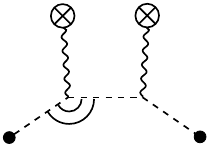}}}+\vcenter{\hbox{\includegraphics[scale=0.4]{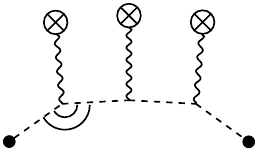}}}+\vcenter{\hbox{\includegraphics[scale=0.4]{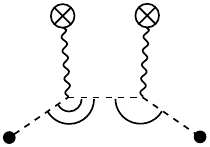}}}+\vcenter{\hbox{\includegraphics[scale=0.4]{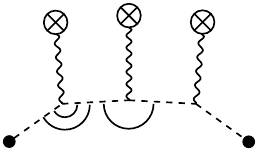}}}\nonumber\\
&+ \vcenter{\hbox{\includegraphics[scale=0.4]{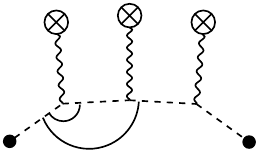}}}+ \vcenter{\hbox{\includegraphics[scale=0.4]{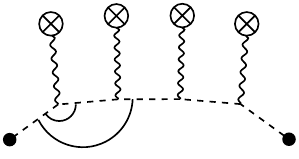}}}+ \vcenter{\hbox{\includegraphics[scale=0.4]{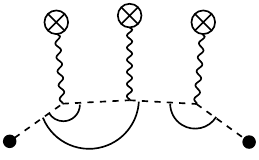}}}+\vcenter{\hbox{\includegraphics[scale=0.4]{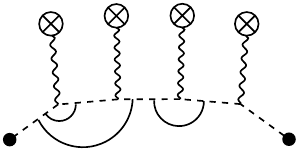}}}.
\end{cases}
\end{align}
\end{widetext}
where $(a)$, $(b)$ and $(c)$ refer to the second, third, and fourth diagrams in the r.h.s of \cref{eqn:nth_dyson} respectively. 
For the $(n+1)$th generation, there will be $m_{n+1} = (m_n+1)^2$ diagrams. Hence the infinite recursion encoded by \cref{eqn:dyson_modified} describes an infinite number of unique contributing diagrams with a `vertex-corrected' topology. While some diagrams appear in both the second and third generation, at each generation there is only one of each diagram, which is also true of $G^{(\infty)}$. Additionally, for this particular theory, no divergent diagrams are generated in the first-order, and therefore no divergent diagrams are generated for subsequent $G^{(n)}$ (because no divergent sub-diagrams are generated). This follows from Dyson's power counting theorem \cite{dyson_s_1949, weinberg_high-energy_1960, zimmermann_power_1968}, and holds if the momentum-space potential decays sufficiently quickly at $\pm \infty$. Because this theory is super-renormalisable anyway, the self-energy terms we neglect remove any divergences. 
The diagrammatic approximation in \cref{eqn:dyson_modified} can be recast into an integral equation
\begin{widetext}
\small
\begin{align}
G(p,k) &\approx \underbrace{-ieD(k,m)D(p,m)\tilde{u}(p-k)}_{\vcenter{\hbox{\includegraphics[scale=0.3]{Figures/chapter_3/feynman_diagrams/ext_interaction_1.pdf}}}}-\underbrace{ieD(p,m)\rint{q}\tilde{u}(p-q)G(q, k)}_{\vcenter{\hbox{\includegraphics[scale=0.3]{Figures/chapter_3/feynman_diagrams/dressed_interaction.pdf}}}}+\underbrace{(-i\mu)^2D(p,m)D(k,m) \rint{q}D(q,m')G(p-q, k-q)}_{\vcenter{\hbox{\includegraphics[scale=0.3]{Figures/chapter_3/feynman_diagrams/dressed_part_self_energy.pdf}}}}\nonumber\\&+\underbrace{(-i\mu)^2D(p,m)\rint{q}\rint{s}D(q,m')G(p-q, s-q)G(s, k)}_{\vcenter{\hbox{\includegraphics[scale=0.3]{Figures/chapter_3/feynman_diagrams/dressed_self_energy.pdf}}}}.
\end{align}
\end{widetext}
\normalsize
We note that the first two terms in the integral equation correspond to the relativistic quantum mechanics propagator, in the absence of the additional field $\Phi$. It is the final two terms which encode QFT corrections. Though approximate, they include all-orders in $\mu$. Though this expression is still complex, significant gains have been achieved: the fact that an integral equation can now encode the explicit corrections to tunnelling gives a way to answer the questions this paper poses. 
 \subsection*{Perturbative coupling approximation \label{sec:perturbative_coupling}}
 Though the external field requires a non-perturbative treatment, there is no reason why the coupling $\mu$ cannot be perturbative, and still yield results compatible with tunnelling. Thus, the next natural approximation is to consider the regime where the
coupling to the external potential is much stronger than the coupling to the scalar field, $\Phi$
(i.e. $eu(x) \gg \mu$). Conceptually, this is the same approximation made in the Furry expansion in strong-field QED \cite{fedotov_advances_2023}: the external electromagnetic field is treated exactly, while the dynamical quantised field is treated perturbatively. This has the effect of removing self-consistency in \cref{eqn:dyson_modified}. To first-order in $\mu^2$, \cref{eqn:dyson_modified} becomes:
\begin{align}
\vcenter{\hbox{\includegraphics[scale=0.7]{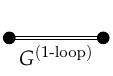}}} =& \vcenter{\hbox{\includegraphics[scale=0.7]{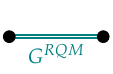}}}+\vcenter{\hbox{\includegraphics[scale=0.7]{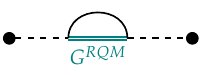}}}+\vcenter{\hbox{\includegraphics[scale=0.7]{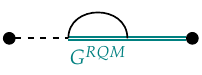}}} \label{eqn:dyson_RQM_approx}\\
=& \vcenter{\hbox{\includegraphics[scale=0.4]{Figures/chapter_3/feynman_diagrams/ext_interaction_1.pdf}}}+ \vcenter{\hbox{\includegraphics[scale=0.4]{Figures/chapter_3/feynman_diagrams/double_external_vertex.pdf}}}+\vcenter{\hbox{\includegraphics[scale=0.4]{Figures/chapter_3/feynman_diagrams/dressed_expansion/second/triple_vertex.pdf}}}+\vcenter{\hbox{\includegraphics[scale=0.4]{Figures/chapter_3/feynman_diagrams/vertex_correction.pdf}}}\nonumber \\
+&\vcenter{\hbox{\includegraphics[scale=0.4]{Figures/chapter_3/feynman_diagrams/dressed_expansion/second/loop_two_leg.pdf}}}
+\vcenter{\hbox{ \includegraphics[scale=0.4]{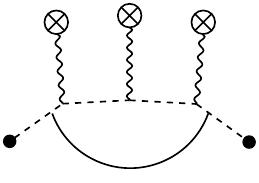}}} \nonumber\\
+& \vcenter{\hbox{\includegraphics[scale=0.4]{Figures/chapter_3/feynman_diagrams/dressed_expansion/first/vertex_correction_leg.pdf}}}+\vcenter{\hbox{\includegraphics[scale=0.4]{Figures/chapter_3/feynman_diagrams/dressed_expansion/second/vertex_correction_double_leg.pdf}}}+\vcenter{\hbox{\includegraphics[scale=0.4]{Figures/chapter_3/feynman_diagrams/dressed_expansion/second/two_point_vertex_one_leg.pdf}}} + \mathcal{O}(e^4) \nonumber
\end{align}
with the relativistic quantum mechanical solution (i.e. the tree-level solution),
\begin{equation}
\vcenter{\hbox{\includegraphics[scale=0.6]{Figures/chapter_3/feynman_diagrams/dressed_line_RQM.pdf}}} = \vcenter{\hbox{\includegraphics[scale=0.55]{Figures/chapter_3/feynman_diagrams/ext_interaction_1.pdf}}}+\vcenter{\hbox{\includegraphics[scale=0.6]{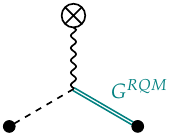}}}.\label{eqn:rqmdyson}\\
\nonumber\\
\end{equation}
However, this overly-simple approximation neglects non-trivial one-loop diagrams. For instance, while \cref{eqn:dyson_modified} necessarily generates the diagrams
\begin{align}
\vcenter{\hbox{\includegraphics[scale=0.5]{Figures/chapter_3/feynman_diagrams/excluded_loop.pdf}}}\quad \text{ and} \quad\vcenter{\hbox{\includegraphics[scale=0.5]{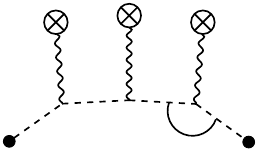}}},
\end{align}
the above insertion of the RQM propagator \textit{does not}, because it is no longer self-referential. To remedy this, we instead redefine the one-loop propagator with additional terms:
\begin{align}
\vcenter{\hbox{ \includegraphics[scale=0.7]{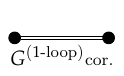}}} =& \vcenter{\hbox{\includegraphics[scale=0.7]{Figures/chapter_3/feynman_diagrams/dressed_line_1.pdf}}}+\vcenter{\hbox{\includegraphics[scale=0.4]{Figures/chapter_3/feynman_diagrams/dressed_expansion/second/vertex_correction_leg_flip.pdf}}}+\vcenter{\hbox{\reflectbox{\includegraphics[scale=0.4]{Figures/chapter_3/feynman_diagrams/dressed_expansion/second/two_point_vertex_one_leg.pdf}}}}\nonumber\\
+&\vcenter{\hbox{\includegraphics[scale=0.4]{Figures/chapter_3/feynman_diagrams/excluded_loop.pdf}}}+\quad\vcenter{\hbox{\includegraphics[scale=0.4]{Figures/chapter_3/feynman_diagrams/excluded_loop_2.pdf}}}+\mathcal{O}(e^4)\nonumber \\
=& \vcenter{\hbox{\includegraphics[scale=0.7]{Figures/chapter_3/feynman_diagrams/dressed_line_1.pdf}}} + \vcenter{\hbox{\includegraphics[scale=0.7]{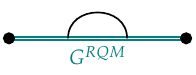}}}+\vcenter{\hbox{\includegraphics[scale=0.7]{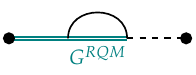}}},\label{eqn:GRQM_expand}
\end{align}
While \cref{eqn:GRQM_expand} is still not self-referential, it does encode all diagrams relevant to tunnelling, at one-loop. It is more concisely expressed via
\begin{equation}
\vcenter{\hbox{ \includegraphics[scale=0.7]{Figures/extra_feynman_diagrams_paper_2/one_loop_correction.pdf}}} = \vcenter{\hbox{\includegraphics[scale=0.7]{Figures/chapter_3/feynman_diagrams/dressed_line_RQM.pdf}}}+\left \{\vcenter{\hbox{\includegraphics[scale=0.7]{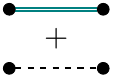}}}\right \}\vcenter{\hbox{\includegraphics[scale=0.7]{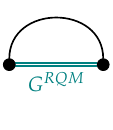}}}\left \{\vcenter{\hbox{\includegraphics[scale=0.7]{Figures/chapter_3/feynman_diagrams/propagator_sum.pdf}}}\right \},\label{eqn:compact_RQM_dyson}
\end{equation}
with the corresponding mathematical expression
\begin{widetext}
\begin{align}
&G^{(1\text{-loop})}(p,k) = G^{RQM}(p,k)+(-i\mu)^2 \rint{q}D(q,m')\Bigg\{\underbrace{
D(p,m)D(k,m)G^{RQM}(p-q, k-q)}_{\vcenter{\hbox{\includegraphics[scale=0.4]{Figures/chapter_3/feynman_diagrams/dressed_part_self_energy_RQM.pdf}}}}
\nonumber\\
&+\underbrace{D(p,m)\rint{s}G^{RQM}(p-q,s-q)G^{RQM}(s,k)}_{\vcenter{\hbox{\includegraphics[scale=0.4]{Figures/chapter_3/feynman_diagrams/dressed_self_energy_RQM.pdf}}}}+\underbrace{D(k,m)\rint{s}G^{RQM}(p, s)G^{RQM}(s-q, k-q)}_{\vcenter{\hbox{\includegraphics[scale=0.4]{Figures/chapter_3/feynman_diagrams/dressed_addition_2.pdf}}}}\nonumber\\
\quad\quad&+\underbrace{\rint{s}\rint{w} G^{RQM}(p,s)G^{RQM}(s-q, w-q)G^{RQM}(w,k)}_{\vcenter{\hbox{\includegraphics[scale=0.4]{Figures/chapter_3/feynman_diagrams/dressed_rqm_addition.pdf}}}}\Bigg \}.\label{eqn:one_loop_RQM_equation}
\end{align}
\end{widetext}
\Cref{eqn:one_loop_RQM_equation} has a vastly simplified structure compared to \cref{eqn:dyson_modified}, given that it is no longer an integral equation. Provided $G^{\text{RQM}}(p,k)$ is known, the numerical determination of the one-loop correction should be feasible. However, finding a form for $G^{\text{RQM}}(p,k)$ is not always straightforward: it is the solution of an integral equation, and moreover, must be non-perturbative in the external field to be useful for tunnelling calculations. 
\subsection*{Discussions and Conclusions}
\Cref{eqn:dyson_modified,eqn:one_loop_RQM_equation} provide the groundwork for future work to quantitatively determine how QFT corrections impact tunnelling probabilities. One promising avenue may be numerical methods, particularly for solving \cref{eqn:one_loop_RQM_equation}, if the form of $G^{\text{RQM}}(p,k)$ is known (or numerically found). We have found analytic expressions for $G^{\text{RQM}}(p,k)$ using techniques of Feynman diagram resummation, for simple potentials such as a Dirac delta and double-delta potential \cite{zielinski_2024_eprint}. While these analytic methods employ calculation techniques in QFT, in principle, $G^{\text{RQM}}$ can be found within a relativistic quantum mechanical framework if needed. This is analogous to how QED corrections are implemented in the Furry expansion - for instance, a recent paper \cite{sommerfeldt_all-order_2023} used the analytic form of the Dirac-Coulomb Green's function to then numerically calculate corrections to Delbr\"uck scattering.

 While Feynman integrals are routinely computed in the context of collider physics, with many \texttt{Mathematica} packages \cite{patel_package-x_2015, patel_package-x_2017, mertig_1991, shtabovenko_2023,smirnov_2022a} devoted to this, \cref{eqn:one_loop_RQM_equation} does not have the mathematical form of a Feynman integral to allow one to exploit these methods. In particular, our tunnelling propagators manifestly lack Lorentz-invariance, which is a property many of these programs require. However, there have been recent developments in the non-relativistic effective field theory space, with modifications to existing packages \cite{brambilla_2020} extending semi-automatic symbolic calculations beyond Lorentz-invariance formulations. This may be a promising avenue for continued work, although currently, it is best suited for tree-level or 1-loop amplitudes. 
Additionally, standard Monte Carlo techniques \cite{filinov_calculation_1986} for computing integrals of our form may encounter difficulty in \cref{eqn:one_loop_RQM_equation}, due to poor convergence from the presence of poles (despite the integrals being formally finite). Numerical contour integration may be a useful avenue, although this still requires a robust method for locating poles in the integrand \cite{capatti_numerical_2020,kermanschah_numerical_2022,becker_direct_2013}. In principle, if this propagator could be numerically evaluated, it would provide enough information to conclusively determine how loop corrections impact tunnelling probabilities. One may then probe the conditions which enhance this effect, and thus how it may be measured. 

The benefit of this work is that the analogy with the self-consistent Hartree-Fock approximation may be readily applied to more physical Lagrangians. For instance, the QED Lagrangian with the addition of a classical vector field would be one candidate, capable of describing corrections to electron tunnelling through electric potentials. A very similar dressed propagator would arise for this theory, given the photon interaction with the electron has a similar cubic structure, and so the same topological diagrams would be retained. However, because this is not a super-renormalisable theory, counter-terms and a relevant renormalisation procedure would be required, as would be the case for most physically interesting Lagrangians.  
\bibliography{QFT_tunnelling_references_master}

\end{document}